\documentstyle[twoside,fleqn,npb,epsfig,axodraw]{article}
\newcommand{\Slash} {\slash \!\!\!}
\title{NNLO Corrections to the Polarized Drell-Yan Coefficient Function.}

\author{V. Ravindran,\address{Harish-Chandra Research Institute, Chhatnag Road, 
Jhusii, Allahabad, 211019, India.}%
        \thanks{Deutsches Elektronen-Synchroton DESY, Platanenallee 6,
15738 Zeuthen.}
        J. Smith,\address{C.N. Yang Institute for Theoretical Physics,
State University of New York at Stony Brook, New York 11794-3840, USA}%
        W.L. van Neerven.\address{Instituut-Lorentz, Universiteit Leiden, 
PO Box 9506, 2300 RA Leiden, The Netherlands.}}

\begin{document}

\begin{abstract}
We present the full next-to-next-to-leading order (NNLO) corrections to
the coefficient function for the polarized cross section $d~\Delta\sigma/d~Q$
of the Drell-Yan process. We study the effect of these corrections
on the process $p+p\rightarrow
l^+l^-+`X'$ at an C.M. energy $\sqrt{S}=200~{\rm GeV}$. All QCD partonic 
subprocesses have been included provided the lepton pair is created by
a virtual photon, which is a valid approximation for a lepton pair invariant 
mass $Q<50~{\rm GeV}$. For this reaction the dominant subprocess is 
given by $q+\bar q\rightarrow \gamma^*+`X'$ and its higher order corrections
so that it provides us with an excellent tool to measure the polarized
sea-quark densities.
\end{abstract}

\maketitle

\section{Introduction}
Deep inelastic electroproduction is very useful to extract information
about the polarized valence parton densities $\Delta u_v$ and $\Delta d_v$. 
However almost no information exists about the gluon density $\Delta g$
and the sea-quark densities $\Delta u_s$, $\Delta d_s$ and $\Delta s$ 
(including the anti-sea-quark densities).
One of the processes proposed to measure the latter densities is the Drell-Yan
process or massive lepton pair production $p+p\rightarrow l^+l^-+`X'$. 
In this reaction the dominant subprocess is given by valence-quark sea-quark 
annihilation into the lepton pair which continues to hold if we include
higher order QCD corrections. Lepton pair production is given by the process
\begin{eqnarray}
\label{eq1.1}
&& p(P_1,S_1)+p(P_2,S_2)\rightarrow \gamma^*(q) + 'X'\,.
\nonumber\\
&& \hspace*{40mm}\mid
\nonumber\\
&& \hspace*{40mm}\rightarrow l^+(l_1)+l^-(l_2)
\nonumber\\[2ex]
&& S=(P_1+P_2)^2 \,, \quad Q^2\equiv q^2 =(l_1+l_2)^2 \,,
\end{eqnarray}
In the frame work of the parton model this process is described by the
Drell-Yan mechanism
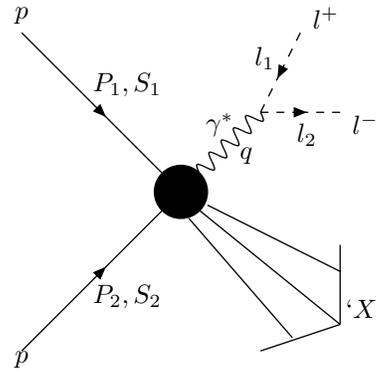
\begin{figure}
\begin{center}
\begin{picture}(130,130)(0,0)
\ArrowLine(0,0)(60,60)
\ArrowLine(0,120)(60,60)
\GCirc(60,60){10}{0}
\Line(67,53)(120,10)
\Line(63,50)(102,5)
\Line(70,55)(120,30)

\Line(90,0)(120,10)
\Line(120,40)(120,10)

\Photon(67,67)(90,90){3}{5}
\DashArrowLine(105,120)(90,90){3}
\DashArrowLine(90,90)(120,90){3}

\Text(0,130)[t]{$p$}
\Text(0,0)[t]{$p$}
\Text(40,25)[t]{$P_2,S_2$}
\Text(40,105)[t]{$P_1,S_1$}
\Text(75,90)[t]{$\gamma^*$}
\Text(85,77)[t]{$q$}
\Text(115,130)[t]{$l^+$}
\Text(130,90)[t]{$l^-$}
\Text(92,115)[t]{$l_1$}
\Text(108,86)[t]{$l_2$}
\Text(130,20)[t]{$`X'$}

\end{picture}
\caption[]{Massive lepton pair production $p+ p
\rightarrow \gamma^* + `X'$. }
\label{fig1}
\end{center}
\end{figure}
where the protons are longitudinally polarized. At $\sqrt{S}=200~{\rm GeV}$
the photon dominates which is characteristic of the RHIC energies at BNL. 
Let us first look at the 
transverse momentum $p_T$ and rapidity $y$ distributions. They are given by
\begin{figure}
\begin{center}
\begin{picture}(130,130)(0,0)
\ArrowLine(0,110)(28,110)
\ArrowLine(120,20)(92,20)
\CArc(45,110)(17,120,240)
\CArc(25,110)(17,300,60)

\CArc(95,20)(17,120,240)
\CArc(75,20)(17,300,60)

\DashArrowLine(40,100)(60,65){3}
\DashArrowLine(80,30)(60,65){3}

\DashArrowLine(42,115)(110,115){3}
\DashArrowLine(42,110)(110,110){3}
\DashArrowLine(42,105)(110,105){3}

\DashArrowLine(78,25)(10,25){3}
\DashArrowLine(78,20)(10,20){3}
\DashArrowLine(78,15)(10,15){3}

\Photon(60,65)(90,65){3}{5}
\ArrowLine(120,75)(90,65)
\ArrowLine(90,65)(120,55)
\Gluon(48,90)(80,90){3}{5}
\Gluon(67,50)(35,50){3}{5}

\Text(0,120)[t]{$p$}
\Text(130,20)[t]{$p$}
\Text(43,85)[t]{$q$}
\Text(78,50)[t]{$\bar q$}
\Text(80,80)[t]{$\gamma^*$}

\Text(85,93)[t]{$g$}
\Text(30,50)[t]{$g$}

\Text(130,80)[t]{$l^+$}
\Text(130,60)[t]{$l^-$}

\end{picture}
\caption[]{The Drell-Yan process for $p+ p
\rightarrow \gamma^* + `X'$. }
\label{fig2}
\end{center}
\end{figure}
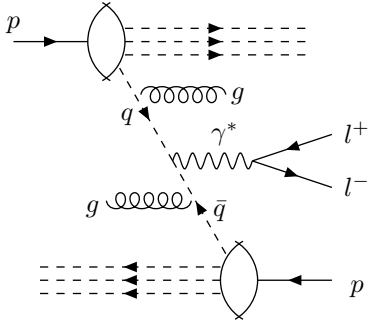
\begin{eqnarray}
S\,\frac{d^3\,\Delta \sigma}{dQ^2\,dp_T\,dy}=\frac{4\,\pi\,\alpha^2}{3\,
N_C\,Q^2}\,\frac{d^2\,\Delta W}{dp_T\,dy}\,,
\end{eqnarray}
where $N_C$ denotes the number of colours.
In the QCD improved parton model we have
\begin{eqnarray}
\frac{d^2\,\Delta W}{dp_T\,dy}&=&\sum_{a_1,a_2=q,g}\int dx_1\,\int dx_2\,
\Delta f_{a_1}(x_1,\mu^2)
\nonumber\\[2ex]
&&\times \Delta f_{a_2}(x_2,\mu^2)\,
\frac{d^2\,\Delta \hat W_{a_1a_2}}{dp_T\,dy} \,,
\end{eqnarray}
where the factorization/renormalization scale is $\mu^2$ and $\Delta f_{a_1}$
is the polarized parton density. The NLO corrections to the partonic 
distribution $d^2\,\Delta \hat W_{a_1a_2}/dp_T\,dy$ have been 
completely calculated in \cite{rsn1} (for the non-singlet part see also 
\cite{chco}). For $p_T>Q/2$ the $O(\alpha_s)$ quark-gluon subprocess dominates 
over all other reactions provided there is a 
substantial gluon density. Therefore a measurement of the 
differential $p_T$ distribution is sensitive to the polarized gluon 
density. This is revealed by the the longitudinal asymmetry 
$A_{LL}=\Delta \sigma/\sigma$ where the set with the largest gluon density
leads to the largest asymmetry.
\begin{figure}[htb]
\centering{\rotatebox{270}{\epsfig{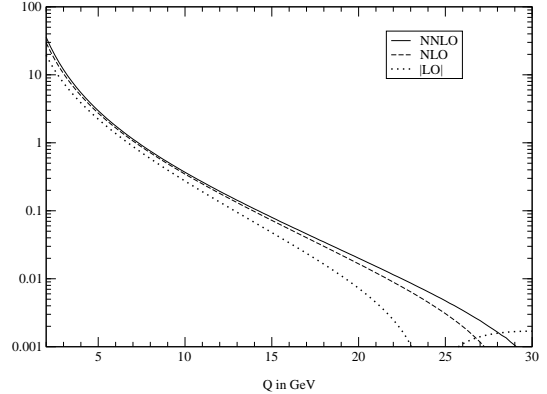}}}
\caption{ The cros section $d\Delta\sigma/dQ$ in pb/GeV plotted in the
range $2 < Q < 30$ GeV at $\mu=Q$ in LO, NLO and NNLO.}
\label{fig:realsmall}
\end{figure}
\begin{figure}[htb]
\vspace{9pt}
\centering{\rotatebox{270}{\epsfig{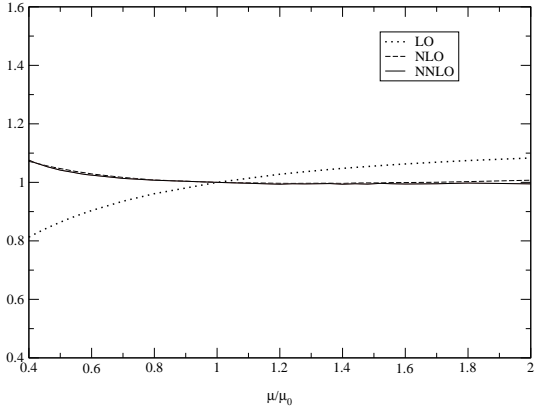}}}
\caption{The polarized quantity $N(\mu/\mu_0)$ plotted in the range
$0.4 < \mu/\mu_0 < 2$ with $Q=5$ GeV and $\mu_0=Q$.}
\label{fig:largenenough}
\end{figure}
\section{Computational details}
A different picture is shown by the invariant mass squared distribution given
by
\begin{eqnarray}
\frac{d\,\Delta\sigma}{dQ^2}=\frac{4\,\pi\,\alpha^2}{3\,N_C\,Q^2\,S}\,
\Delta W\left (\tau,\frac{Q^2}{\mu^2}\right )\,,\quad \tau=\frac{Q^2}{S}\,.
\end{eqnarray}
The structure function $\Delta W$ equals
\begin{eqnarray}
&&\Delta W\left (\tau,\frac{Q^2}{\mu^2}\right )=\sum_{a_1,a_2=q,g}
\int_{\tau}^1 dx_1\,\int_{\tau/x_1}^1 dx_2\,
\nonumber\\[2ex]
&&\times \Delta f_{a_1}(x_1,\mu^2)\, \Delta f_{a_2}(x_2,\mu^2)\,
\nonumber\\[2ex]
&&\times \Delta_{a_1a_2}\left (\tau,\frac{Q^2}{\mu^2}\right )\,.
\end{eqnarray}
Here the zeroth order quark-anti-quark process dominates so that this process
is an excellent tool to measure the sea-quark density in proton-proton 
reactions.
The NLO corrections to the coefficient function $\Delta_{a_1a_2}$ were 
calculated in \cite{rat} and the NNLO corrections have been recently
computed in \cite{rsn2}. In higher order the quark-anti-quark process
also dominates. If $\tilde M$ is the amplitude of a parton-parton subprocess
then the $q\bar q$ is given by
\begin{eqnarray}
\label{eqn2.9}
q(p_1,s_1)+\bar q(p_2,s_2) \rightarrow \gamma^* + 'X'\,,
\nonumber
\end{eqnarray}
\begin{eqnarray}
|\Delta M_{q\bar q}|^2&=&\frac{1}{4}\,{\bf Tr}\,\left (\gamma_5{\Slash}s_2\,
({\Slash}p_2-m) \,\tilde M \,\gamma_5{\Slash}s_1 \right.
\nonumber\\[2ex]
&& \left. \times ({\Slash}p_1+m)\tilde M^{\dagger}\right )\,,
\end{eqnarray}
\begin{eqnarray}
\label{eqn2.10}
q_1(\bar q_1)(p_1,s_1)+ g(p_2) \rightarrow \gamma^* + 'X'\,,
\nonumber
\end{eqnarray}
\begin{eqnarray}
|\Delta M_{qg}|^2&=&\frac{1}{4}\,\epsilon_{\mu\nu\lambda\sigma}\,
\frac{p_2^{\lambda} \,l_2^{\sigma}}{p_2\cdot l_2}\,{\bf Tr}\,\left
(\tilde M^{\mu}
\,\gamma_5{\Slash}s_1 \right.
\nonumber\\[2ex]
&& \left. \times ({\Slash}p_1\pm m)\, \tilde M^{\nu\dagger}\right )\,,
\end{eqnarray}
\begin{eqnarray}
q_1(\bar q_1)(p_1,s_1)+ q_2(\bar q_2)(p_2,s_2) \rightarrow \gamma^* + 'X'
\nonumber
\end{eqnarray}
\begin{eqnarray}
|\Delta M_{q_1q_2}|^2&=&\frac{1}{4}\,{\bf Tr}\,\left (\gamma_5{\Slash}s_2\,
({\Slash}p_2\pm m) \,\tilde M \,\gamma_5{\Slash}s_1 \right.
\nonumber\\[2ex]
&& \left. \times({\Slash}p_1\pm m)\, \tilde M^{\dagger} \right )\,,
\end{eqnarray}
\begin{eqnarray}
 g(p_1)+ g(p_2) \rightarrow \gamma^* + 'X'\,,
\nonumber
\end{eqnarray}
\begin{eqnarray}
|\Delta M_{gg}|^2&=&
\frac{1}{4}\,\epsilon_{\mu_2\nu_2\lambda_2\sigma_2}\,
\frac{p_2^{\lambda_2} \,l_2^{\sigma_2}}{p_2\cdot l_2}\,
\epsilon_{\mu_1\nu_1\lambda_1\sigma_1}
\nonumber\\[2ex]
&&\times \frac{p_1^{\lambda_1}
\,l_1^{\sigma_1}}{p_1\cdot l_1}\,{\bf Tr}\,\left (\tilde M^{\mu_1\mu_2}\,
\tilde M^{\nu_1\nu_2\dagger}\right )\,,
\end{eqnarray}
with the constraints
\begin{eqnarray}
s_i\cdot p_i =0\,,\quad s_i\cdot s_i=-1 \,,\quad l_i\cdot l_i=0\,.
\end{eqnarray}
All these reactions contain 
the $\gamma_5$-matrix and/or the Levi-Civita tensor. 
If we use $n$-dimensional regularization we have to find a prescription
for these typical four dimensional objects. Here we choose the prescription
of HVBM \cite{hvbm} or equivalently the one given by \cite{akde}. Choosing
the latter we obtain the prescription
\begin{itemize}
\item[1.]
Replace the $\gamma_5$-matrix by
\begin{eqnarray}
\gamma_{\mu}\,\gamma_5&=&\frac{i}{6}\,\epsilon_{\mu\rho\sigma\tau}\,
\gamma^{\rho}\, \gamma^{\sigma}\,\gamma^{\tau}\quad \mbox {or} 
\nonumber\\[2ex]
\gamma_5&=&\frac{i}{24}\,\epsilon_{\rho\sigma\tau\kappa}\,\gamma^{\rho}\,
\gamma^{\sigma}\,\gamma^{\tau}\,\gamma^{\kappa} \,.
\nonumber
\end{eqnarray}
\item[2.]
Compute all matrix elements in $n$ dimensions.
\item[3.]
Evaluate all Feynman integrals and phase space integrals in $n$-dimensions.
\item[4.]
Contract the Levi-Civita tensors in four dimensions after the Feynman
integrals and phase space integrals are carried out.
\end{itemize}
This procedure requires an intensive tensorial reduction of the loop-integrals
as well as the phase space integrals. However the HVBM methods entails
the presence of evanescent counter terms. They were calculated up to two-loop
order in \cite{masm}. The result has the following structure
\begin{eqnarray}
\label{eq2.15}
Z_{qq}^{5,{\rm NS},+}&=&\delta(1-x)+a_s\,S_\varepsilon\,
\left (\frac{Q^2}{\mu^2}\right )^{\varepsilon/2}\,\Bigg [z^{(1)}_{qq}
\nonumber\\[2ex]
&&+\varepsilon \bar z^{(1)}_{qq}\Bigg ]
+\hat a_s^2\,S_\varepsilon^2\,\left (\frac
{Q^2}{\mu^2}\right )^\varepsilon
\nonumber\\[2ex]
&&\times\Bigg[
+\frac{1}{\varepsilon}\,\beta_0\,z^{(1)}_{qq}+ z^{(2),{\rm NS},+}_{qq}
\Bigg ]\,,
\end{eqnarray}
where $a_s=\alpha_s(\mu^2)/4\pi$ is the renormalized coupling constant.
Further $\bar z^{(1)}_{qq}$ and $(Q^2/\mu^2)^{k\varepsilon}$ are process
dependent. The other constants are process independent. The latter also
holds for the following pieces
\begin{eqnarray}
\label{eq2.16}
Z_{qq}^{5,{\rm NS},-}=-\hat a_s^2\,S_\varepsilon^2\,\left (\frac{Q^2}{\mu^2}
\right )^\varepsilon \,\Bigg [z^{(2),{\rm NS},-}_{qq} \Bigg ]\,,
\end{eqnarray}
\begin{eqnarray}
\label{eq2.17}
Z_{qq}^{5,{\rm PS}}=\hat a_s^2\,S_\varepsilon^2\,\left (\frac{Q^2}{\mu^2}
\right )^\varepsilon \,\Bigg [z^{(2),{\rm PS}}_{qq}\Bigg ] \,,
\end{eqnarray}
The constant $Z_{qq}^{5,{\rm NS},+}$ is chosen in such a way that
the following relation holds
\begin{eqnarray}
\label{2.18}
\left (Z_{qq}^{5,{\rm NS},+}\right )^2=-\frac{\Delta \hat W_{q\bar q}^{\rm NS}
(\hat a_s,Q^2/\mu^2)}{\hat W_{q\bar q}^{\rm NS}(\hat a_s,Q^2/\mu^2)}
\end{eqnarray}
where $\hat W_{q\bar q}^{\rm NS}$ is the partonic structure function
of the unpolarized reaction. We have to remove the spurious terms coming
from the $\gamma_5$ and Levi-Civita prescription in the partonic structure 
function by using these evanescent counter terms. Since they are present
in the (anti-)quark sector we have to form the following products 
\begin{eqnarray}
&&\left (Z_{qq}^{5}\right )^{-2}\,\Delta W_{q\bar q}\,,\quad 
\left (Z_{qq}^{5}\right )^{-1}\,\Delta W_{qg}\,,\quad
\nonumber\\[2ex]
&&\left (Z_{qq}^{5}\right )^{-2}\,\Delta W_{qq}\,,\quad\Delta W_{gg}
\end{eqnarray}
\begin{figure}[htb]
\centering{\rotatebox{270}{\epsfig{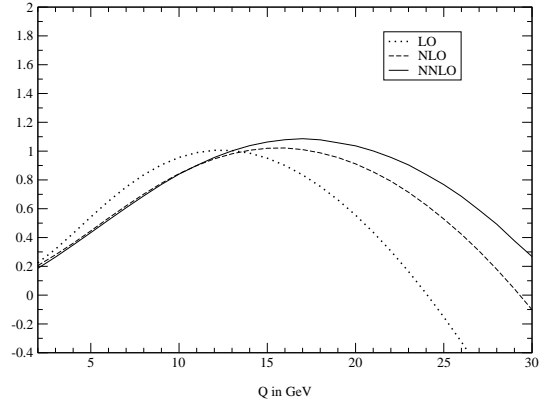}}}
\caption{The longitudinal asymmetry $A_{LL}$ plotted in percent in the range
$2 < Q < 30$ GeV at $\mu=Q$ in LO, NLO and NNLO.}
\label{fig:toosmall}
\end{figure}
\section{Results}
The processes which have to be calculated up to NNLO are
\begin{eqnarray}
\label{eq3.1}
q + \bar q \rightarrow \gamma^*\,.
\end{eqnarray}
\begin{eqnarray}
\label{eq3.2}
q + \bar q \rightarrow \gamma^* + g\,,
\end{eqnarray}
\begin{eqnarray}
\label{eq3.3}
g + q(\bar q) \rightarrow \gamma^* + q(\bar q)\,.
\end{eqnarray}
\begin{eqnarray}
\label{eq3.4}
q + \bar q \rightarrow \gamma^* + g + g\,,
\end{eqnarray}
\begin{eqnarray}
\label{eq3.5}
g + q(\bar q) \rightarrow \gamma^* + q(\bar q) + g\,,
\end{eqnarray}
\begin{eqnarray}
\label{eq3.6}
q + \bar q \rightarrow \gamma^* + q + \bar q \,,
\end{eqnarray}
\begin{eqnarray}
\label{eq3.7}
q (\bar q) + q (\bar q) \rightarrow \gamma^* + q (\bar q) + q(\bar q)\,,
\end{eqnarray}
\begin{eqnarray}
\label{eq3.8}
g + g \rightarrow \gamma^* + q + \bar q\,.
\end{eqnarray}
including virtual corrections to lower order processes. Apart from
evanescent counter terms which are characteristic of the $\gamma_5$
and Levi-Civita prescription the calculation proceeds in the same
way as in the unpolarized case. The renormalization and mass factorization 
are carried out in the ${\overline {MS}}$-scheme. We use the BB1 \cite{blbo}
parton density set to make our plots. The figures are constructed as follows
\begin{tabbing}
aaaaaaaaaaaaa \= aaaaaaaaaaaaaaaaaa \= aaaaaaaaaaaaaaaaaaaa  \kill
$LO$ \>  $\Delta f^{LO}(x,\mu^2)$ \> $\alpha_s^{LO}$\\
        \>                 \>                \\
$NLO$ \> $\Delta f^{NLO}(x,\mu^2)$ \> $\alpha_s^{NLO}$\\
        \>                    \>               \\
$NNLO$\>   $\Delta f^{NLO}(x,\mu^2)$\>  $\alpha_s^{NLO}$
\end{tabbing}
Because of the lack of the three-loop anomalous dimensions we have no
$\Delta f^{NNLO}$. Therefore in NNLO we use the NLO parton density set
and the two-loop corrected running coupling constant $\alpha_s^{NLO}$.
Further we set the mass factorization scale equal to the renormalization 
scale unless stated otherwise.
Since we also compare with the unpolarized Drell-Yan process we choose
the following parton density sets given by \cite{maro}. Since they have an 
approximate NNLO set we plot the figures according to
\begin{tabbing}
aaaaaaaaaaaaa \= aaaaaaaaaaaaaaaaaa \= aaaaaaaaaaaaaaaaaaaa  \kill
$LO$ \>  $f^{LO}(x,\mu^2)$ \> $\alpha_s^{LO}$\\
        \>                 \>                \\
$NLO$ \> $f^{NLO}(x,\mu^2)$ \> $\alpha_s^{NLO}$\\
        \>                    \>               \\
$NNLO$\>   $f^{NNLO}(x,\mu^2)$\>  $\alpha_s^{NNLO}$
\end{tabbing}
Our findings are that the quark-anti-quark channel dominates the 
proton-proton reaction in polarized as well as well in unpolarized physics.
The invariant mass distribution is shown in Fig. 3 in LO, NLO
and NNLO. The K-factors defined by
\begin{eqnarray}
K^{(i)}=\frac{\sigma^{(i)}}{\sigma^{LO}} \,, \qquad 
\Delta K^{(i)}=\frac{\Delta\,\sigma^{(i)}}{\Delta\,\sigma^{LO}}\,,
\end{eqnarray}
are roughly the same. For $Q=7~{\rm GeV}$ at $\sqrt{S}=200~{\rm GeV}$
we get a minimum. Here the values are $\Delta K^{NLO}=1.2$
and $\Delta K^{NNLO}=1.3$ respectively. In Fig. 4 we have plotted for
$Q=5~{\rm GeV}$ the mass factorization scale dependence 
\begin{eqnarray}
N\left (\frac{\mu}{\mu_0}\right )=\frac{\Delta \sigma(\mu)}
{\Delta \sigma(\mu_0)}\,,
\end{eqnarray}
for $\mu_0=Q$ and $0.4\,\mu_0<\mu< 2\,\mu_0$. We see a clear improvement
while going from LO to NLO and surprisingly also from NLO to NNLO. In 
Fig. 5 we have shown the longitudinal asymmetry
\begin{eqnarray}
A_{LL}=\frac{\Delta\,\sigma}{\sigma}\,,
\end{eqnarray}
in LO, NLO and NNLO.
To compare the different polarized parton densities
we have also plotted in Fig.6 the longitudinal asymmetry for the 
BB2 \cite{blbo} and the GRSV standard (SS) and valence (VS) parametrizations 
\cite{grsv}. The figure reveals that the largest gluon (BB1) 
leads to smallest
asymmetry. On the other hand the smallest gluon (VS) gives the largest
asymmetry. Notice that the latter parametrization produces the smallest
asymmetry in the differential $p_T$-distribution.
\begin{figure}[htb]
\centering{\rotatebox{270}{\epsfig{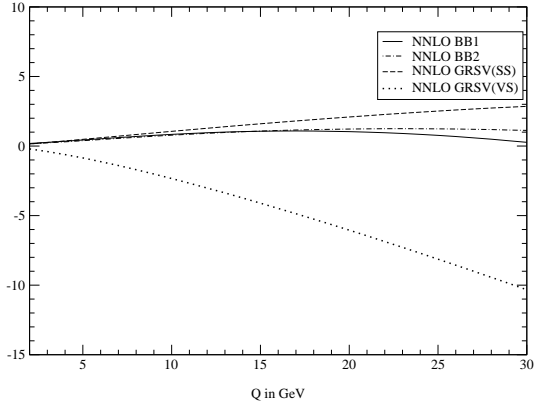}}}
\caption{The longitudinal asymmetry $A_{LL}$ plotted in percent in the range
$2 < Q < 30$ GeV at $\mu=Q$ and NNLO.}
\label{fig:notsmall}
\end{figure}

\end{document}